\documentclass[aps,prl,twocolumn,floatfix]{revtex4-1}

\usepackage{amssymb}
\usepackage{amsmath,amsfonts}
\usepackage{latexsym}
\usepackage{graphicx} 

\def\be{\begin{equation}} 
\def\ee{\end{equation}} 
\def\ba{\begin{eqnarray}} 
\def\ea{\end{eqnarray}}

\def\nl{\nonumber \\}

\def\wh{\widehat}

\def\a{\alpha}

\def\f{\phi}

\def\winf{W_{1+\infty}} 
\def\u1{\widehat{U(1)}}
\def\su2{\widehat{SU(2)}_1}

\begin{document}


\title{Composite Fermion Wavefunctions Derived by Conformal Field Theory}

\author{Andrea Cappelli}
\email{andrea.cappelli@fi.infn.it}

\affiliation{INFN, Via G. Sansone 1, 50019 Sesto Fiorentino - Firenze, Italy}

\date{\today}

\begin{abstract}
The Jain theory of hierarchical Hall states is reconsidered in the
light of recent analyses that have found exact relations between
projected Jain wavefunctions and conformal field theory correlators.
We show that the underlying conformal theory is precisely given by the
W-infinity minimal models introduced earlier. This theory involves a
reduction of the multicomponent Abelian theory that is similar to the
projection to the lowest Landau level in the Jain approach.  The
projection yields quasihole excitations obeying non-Abelian fractional
statistics.  The analysis closely parallels the bosonic conformal
theory description of the Pfaffian and Read-Rezayi states.
\end{abstract}


\maketitle

{\it Introduction.---}
The search for fractional exchange statistics in quasiparticle excitations
of the quantum Hall effect is actively pursued both theoretically
\cite{na-interf}\cite{thermop} and 
experimentally \cite{cb-exp}\cite{thermo-exp}.
The observation of non-Abelian, i.e. multidimensional, statistics,
first suggested in the Pfaffian state \cite{pfaff}, would be
particularly interesting as it could find application 
to quantum computation \cite{tqc}.

The methods of two-dimensional conformal field theory (CFT) \cite{cft}
have been extremely useful in this domain.  
Their description is twofold: they model the dynamics of massless
chiral excitations at the edge of the Hall droplet \cite{wen}, and
express analytic wavefunctions for electrons in the lowest Landau
level.
In particular, the CFT wavefunctions include non-analytic prefactors
that make explicit the fractional statistics of excitations.  All
these methods have been fully developed in the case of Pfaffian
and Read-Rezayi \cite{rr} states, where they have been
instrumental for obtaining several physical results.

The studies of hierarchical Hall states have somehow remained behind,
in spite of their experimental relevance stemming from their richness
and higher stability.  On one side, the Jain theory of the composite
fermion \cite{jain} predicts very accurate ground-state wavefunctions
and has been confirmed by many experiments, but it does not provide
full understanding of quasiparticle statistics.  On the other side,
the CFT description was based on the multicomponent Abelian theory
(multicomponent Luttinger liquid) \cite{hiera}.  This approach has had some
problems, such as the prediction of several distinct types of
electrons.

The precise relation between these two approaches has remained rather
unclear till recent results that form the basis of the present work.
In a series of papers, Hansson et al \cite{hansson} have found 
{\it exact} CFT expressions for the Jain wavefunctions projected to the
lowest Landau level.  These functions have been written in terms of
the expected multicomponent Abelian CFT plus some additional
requirements to be fully interpreted.  Other works \cite{gaffnian}
have studied the short-distance and pairing (clustering) properties of
Jain wavefunctions and compared them to those of the Pfaffian and
other non-Abelian Hall states.

In this Letter, we shall relate these result with another theory of
the hierarchical Hall states that has been formulated independently 
of the composite fermion picture, by Trugenberger, Zemba 
and the author \cite{w-inf}. 
Its main physical input is the incompressibility of the electron fluid
and its symmetry under {\it area-preserving diffeomorphisms} of the
plane, also called W-infinity ($\winf$) symmetry.
The implementation of this symmetry in the dynamics of edge
excitations and the use of results of representation theory
lead to a general analysis of conformal theories suitable for spin-polarized
Hall states.
While they are generally found to be equivalent to the
multicomponent Abelian theories, there are special cases with enhanced
 $SU(n)$ symmetry, where a projection of degrees of freedom is needed to obtain
irreducible representations, i.e. elementary excitations. This is
achieved through a coset construction \cite{cft}, as follows: 
\be
\wh{U(1)}^n \ \longrightarrow \ \wh{U(1)} \times \wh{SU(n)}_1
\ \longrightarrow \ \wh{U(1)} \times \frac{\wh{SU(n)}_1}{SU(n)}\ .
\label{w-coset}\ee
The conformal theories obtained in this way were called $\winf$
minimal models \cite{w-min}; they were found to be in one-to-one
correspondence with the Jain states, but to differ from the standard
multicomponent Abelian theory owing to the projection of the $SU(n)$
symmetry which, among other things, leads to a unique type of electron
and to non-Abelian fusion rules. This projection was also described by
introducing a term in the edge Hamiltonian \cite{w-ham}.  However,
this theory was not yet applied to describe wavefunctions: in this
Letter, we complete the analysis and establish a direct relation with
the Jain composite fermion theory.

{\it Results.---} (i) We show that the Hansson et al. expression of
Jain wavefunctions can be completely {\it derived} within the $\winf$
minimal models.  Furthermore, the projection of the multicomponent
Abelian theory realized in the minimal models can be related with the
Jain projection to the lowest Landau level.

(ii) We remark that the Jain wavefunctions in the Hansson et al. form
are remarkably similar to the Pfaffian and Read-Rezayi wavefunctions
expressed in terms of another projection of Abelian theories 
\cite{cgt}; the analogies between the two approaches are
emphasized.

(iii) We argue that quasiholes excitations over the Jain states should
possess non-Abelian fractional statistics as predicted by the $\winf$
minimal models; the mechanism for non-Abelian statistics is the same as
that of the Pfaffian state and results from the identification of
the independent components (effective Landau levels) down to a single one.

{\it Hansson et al. results.---}
The Jain theory is based~on a correspondence between integer, 
$1/\nu^*=1/n$, and fractional, $1/\nu= p + 1/n$, Hall
effects ($p$ even); this is realized by introducing the 
ground state wave function \cite{jain}:
\be
\Psi_\nu = {\cal P}_{LLL}\ 
\prod_{i<j}^N   \left( z_i -z_j \right)^p \ 
\Psi_{\nu^*=n} \ .
\label{jain-wf}\ee
The term $\Psi_{\nu^*=n} $ is the Slater determinant for $N$ electrons
completely filling the first $n$ Landau levels, say putting $N/n$ of
them in each level; ${\cal P}_{LLL}$ expresses 
the projection to the lowest Landau level.
Note that electrons originally placed in the $j$-th level, $j=1,2,\dots$,
have angular momentum of one-particle states shifted by $(1-j)$.

In their careful study of quasi-particle excitations over the Laughlin
and other states, Hansson et al.  \cite{hansson} obtained an exact
rewriting of (\ref{jain-wf}). Using non-trivial determinant
identities, they found the following formula:
\be
\Psi_\nu = {\cal A}\left[
\prod^{N/2}_{i<j} w_{ij}^{p+1} \ 
\prod^{N/2}_{k=1}\partial_{z_k}
\prod^{N/2}_{i<j} z_{ij}^{p+1} \prod^{N/2}_{i,j} (z_i-w_j)^p  
\right].
\label{hans-wf}\ee
In this expression, we considered the $n=2$ case for simplicity
($\nu=2/(2p+1)$), and put $N_1=N_2=N/2$ electrons
in the first and second level, respectively 
(the result can be extended to $N_1> N_2$ and also to
$n>2$ \cite{hansson}). Electron coordinates in the first
and second level were denoted by $w_i$ and $z_j$, $i,j=1,\dots,N/2$, 
respectively, and their differences by $w_{ij}=w_i -w_j$ and
$z_{ij}=z_i-z_j$. The symbol ${\cal A}$ indicates antisymmetrization with 
respect to all $N$ electron coordinates.

The remarkable expression (\ref{hans-wf}) admits
a simple description in terms of CFT correlators. From the 
 Jastrow factors inside square brakets, one can recognize
the wavefunction of the standard two-component Abelian CFT description
 of  hierarchical states \cite{hiera} with so-called $K$ matrix equal to
$ \bigl(\begin{smallmatrix}
p+1 & p \\ p & p+1
\end{smallmatrix} \bigr)$;
this corresponds to the following correlator of two kinds 
of conformal fields (vertex operators):
\be
\Psi_\nu =
{\cal A}\left[
\left\langle
\left(\partial_{z_1} V_+ \right)\cdots
\left( \partial_{z_{N/2}} V_+\right) V_- \cdots V_-
\right\rangle\right].
\label{conf-wf}\ee
The vertex operators can be written
in terms of charged $\f$ and neutral $\varphi$  scalar fields as
$V_{\pm} = e^{i\sqrt{p+\frac{1}{2}}\, \f}
e^{\pm i\frac{1}{\sqrt{2}} \varphi} $
(see \cite{w-min} for details).

The expressions (\ref{hans-wf}) and (\ref{conf-wf}) 
can be interpreted as describing electrons
belonging to the first two Landau levels, with the derivatives realizing
the shift of angular momentum for the second level
and the antisymmetrization ensuring identical electrons.


{\it Analogies with the Pfaffian state.---} Hansson et al. expression
(\ref{hans-wf}) is very similar to the Pfaffian state in the
two-component Abelian CFT description developed in \cite{cgt}, which
reads ($1/\nu=M+1$, $M$ odd): 
\be 
\Psi_{\rm Pfaff} = {\cal A}\left[
  \prod^{N/2}_{i<j} w_{ij}^{M+2} \ \prod^{N/2}_{i<j} z_{ij}^{M+2}
  \prod^{N/2}_{i,j} (z_i-w_j)^M \right],
\label{pfaff-wf}\ee
corresponding to the matrix
$K= \bigl(\begin{smallmatrix}
M+2 & M \\ M & M+2
\end{smallmatrix} \bigr)$.
After antisymmetrization, the expression (\ref{pfaff-wf}) 
can be shown to be identical to the standard form 
$\Psi_{\rm Pfaff}= \prod z_{ij}^{M+1} \, {\rm Pf}\left(1/z_{ij}\right)$
\cite{pfaff}.

The explanation for such Abelian representation of the Pfaffian
state rests in the physics of electron pairing. Before
antisymmetrization, the state (\ref{pfaff-wf}) describes distinguishable
electrons, possessing a twofold quantum number, say isospin. 
Their wave function does not vanish (for $M=0$) when two electrons
with different isospin meet at the same point $z_1=z_2=z$, indicating
their pairing; when a third electron approaches, it necessarily vanish
as $\Psi\sim (z_3-z)^2$.  After (anti)symmetrization, all electrons
become identical, i.e. the theory only contains isospin singlets, but
the pairing property is retained, that is characteristic of the
Pfaffian state.

The Abelian representation of the Pfaffian 
extends easily to quasihole excitations \cite{cgt}: before projection, these
are of two kinds, each one coupling to its electron: after projection,
they become identical and yield a multidimensional representation
of the braid group, i.e. to non-Abelian statistics \cite{nw}.
This is a nice way to understand non-Abelian statistics
within the standard Abelian setting: it is just a consequence of
the projection to identical electrons.

The similarity of the expressions (\ref{hans-wf}) and (\ref{pfaff-wf})
for the Jain and Pfaffian states brings in a series of results
that will be relevant for the following discussion.
A first observation is that the Jain states possess the same vanishing
behaviour of the Pfaffian and Read-Rezayi states. 
When three or more particles approach
the same point, Eq. (\ref{hans-wf}) behaves as:
\ba
\!\!\!\Psi_{n=2} &\sim& z_{12}^{p-1}\left(
z_{13}^{p+1}\ z_{14}^{p+1} \cdots \right),
\qquad\quad\quad
 \frac{1}{\nu}=p+\frac{1}{2},
\label{clust2}\\
\!\!\!\Psi_{n=3} &\sim& 
\left(z_{12}\, z_{13}\, z_{23}\right)^{p-1}
\left(z_{14}\ z_{15} \cdots \right)^{p+1},
\frac{1}{\nu}=p+\frac{1}{3}.
\label{clust3}\ea
In the second expression, we also reported the $n=3$ case to
be compared with the $\mathbb{Z}_3$ parafermionic Read-Rezayi state
\cite{rr}\cite{cgt}. The result (\ref{clust2}) can be easily proven by
using a graphical representation for the action of derivatives in the
expression (\ref{hans-wf}) \cite{long}.

Therefore, we see that the $M=0$ Pfaffian and the $p=1$ Jain states
have the same pairing properties.  The difference between them is that
the Pfaffian (Read-Rezayi) state is the lowest order polynomial,
i.e. lowest angular momentum state, obeying that type of pairing
(clustering), while the Jain state has higher angular momentum.
Another way to put this fact is that the Pfaffian is the lowest
momentum zero-energy state of a three-body short-distance potential,
while Jain state is an excited state. A short-distance potential that
could similarly single out the Jain state has not been found
\cite{gaffnian}.

{\it Derivation of Jain wavefunctions.---} 
We start from the two-component Abelian theory and describe the
projection of the $SU(2)$ symmetry (\ref{w-coset}) on wavefunctions
(see \cite{w-ham} for an introduction).  The neutral
parts of the vertex operators $V_\pm$ in (\ref{conf-wf}) have
associated isospins values $t_z=\pm 1/2$ that should be identified;
the resulting field is only characterized by its Virasoro dimension
$h=t^2/2$, according to the coset projection (\ref{w-coset}):
\be
\wh{U(1)}  \times 
\frac{\wh{SU(2)}_1}{SU(2)}\sim \wh{U(1)}  \times {\rm Vir}\ .
\label{w-coset2}\ee
A convenient way to perform the projection is by using the Dotsenko-Fateev
screening operators $Q_\pm$ \cite{cft}. These have vanishing scale dimension
but non-vanishing isospin $t_z=\pm 1$, and can  
relate the two electron fields:
\be
V_- \sim V_+ =Q_+ V_-, 
\quad
Q_+ = J^+_0=\oint\! du\, J^+(u)\ .
\label{q-op}\ee
Note that in this theory the screening operators are the zero modes of the
${\wh{SU(2)}_1}$ affine algebra. Using a standard procedure, the
correlator of, say, four electrons can be expressed in terms of four
vertex operators and two screening charges, needed for achieving 
zero total isospin, as follows:
\be
\Psi= \left\langle
\oint_{C_1}\!\!\! J^+ \oint_{C_2}\!\!\! J^+\
V_-(z_1) V_-(z_2) V_-(z_3) V_-(z_4) 
\right\rangle .
\label{4p-wf}\ee

The next step is to choose the contours $C_1,C_2$. In principle,
there are several choices, corresponding to different intermediate states
in the fusion of two electrons. Actually, after projection the
fields in the Virasoro theory acquire non-Abelian fusion rules,
corresponding to the addition of isospins, 
$\{1/2\}\times \{1/2\}=\{0\}+\{1\}$ (these are nothing else than the
the $c\to 1$ limit of the fusion rules of minimal Virasoro models).

\begin{figure}[t]
\includegraphics[width=6cm]{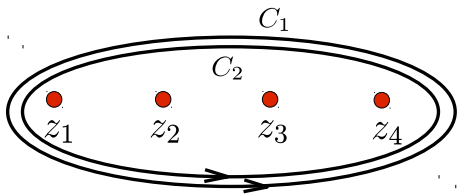}
\caption{The contours for the screening charges.}
\label{fig}
\end{figure}

The ground state wavefunction should be 
completely (anti)symmetric with respect to all electrons; this
requires the contours to encircle all of them as shown in
Fig. \ref{fig}.  Unfortunately, this choice yields a vanishing result
(explicitly checked in \cite{w-ham}), because the contours can be
deformed at infinity where they correspond to acting on the invariant
right vacuum, $\langle 0 | J^-_0=0$.

The solution of this puzzle is to identify the two electron fields
through another ${\wh{SU(2)}_1}$ generator with same $t_z=1$ isospin but
leading to non-vanishing results. The natural replacement is,
\be
J^+_0 \ \to\ J^+_{-1} .
\label{j-shift}\ee
Acting on vertex operators, $J^+_{-1}$ yields the first descendent 
in the tower of states of the affine representation,
\be
J^+_{-1} V_- \sim \partial_z V_+\ ,
\label{j-der}\ee
as easily checked by using the vertex operator algebra.
Note that the operators
$\{J^+_{-1},J^-_1,1/2-J^3_0 \}$ form another $SU(2)$ algebra contained
in the affine algebra ${\wh{SU(2)}_1}$ \cite{cft}, that is equivalent for
realizing the projection (\ref{w-coset2}). 

Thus, we consider the following expression for the four electron
wavefunction:
\be
\Psi'= \left\langle
J^+_{-1}\! V_-(z_1)\,
J^+_{-1}\! V_-(z_2) \, V_-(z_3)V_-(z_4) 
\right\rangle + {\rm perm}.
\label{4p-wf2}\ee
This is actually equal to the Hansson et al. expression (\ref{conf-wf}), 
owing to (\ref{j-der}).

Therefore, we have obtained a derivation of the Jain composite fermion
wavefunction entirely from symmetry arguments, through the
construction of the $\winf$ minimal models and their application to
describe wavefunctions. This results is rewarding insofar as it supports
the universality and robustness of the Jain theory; furthermore, it
can be used to obtain a consistent description of excitations, as
discussed in the next subsection.

Let us remark that Eq.(\ref{j-der}) strictly holds for the neutral part
of the vertex operators $V_\pm$: for the derivative to act on the whole
electron field, as in (\ref{conf-wf}), one needs to deform the $SU(2)$
generator for including a charged $\wh{U(1)}$ part:
$J^+_{-1}\to J^+_{-1} +\sqrt{2}\, \a_0 \a_{-1}$. Upon varying the
relative coefficient, it also possible to obtain a $p$-dependent family 
of wave functions, the Jain wavefunction being one of them. 
This ambiguity could be fixed by other physical requirements,
but the pairing (\ref{clust2}) and fractional statistics properties
are independent of it.
Note also that the $\winf$ derivation of Jain wavefunctions extends
to more than two Landau levels \cite{long}: for $n=3$, two
$SU(3)$ generators $J_{-1}^{\a_i}$ are employed, whose action on vertex
operators leads to single and double derivatives in agreement with 
Hansson et al. results \cite{hansson}.

{\it Quasihole excitations and non-Abelian statistics.---} 
Here we shall be rather brief for lack of space, and defer to
\cite{long} for a complete discussion. The analysis is similar to 
that of the Pfaffian state \cite{nw} in its Abelian description
\cite{cgt}.  In the two-component Abelian theory of Jain states before
projection, there are two quasihole excitations of smallest charge
with isospins $t_z=\pm 1/2$ \cite{w-ham}.  They are represented by
fields $H_\pm$ that couple to the respective electron fields $V_\pm$,
owing to the fusion rules $H_\pm V_\mp\sim I$.  In the Jain
description, each quasihole is made in one effective Landau level, and
Hansson et al. have verified that the insertion of $H_\pm$ in the CFT
correlator (\ref{conf-wf}) reproduces the corresponding Jain
wavefunction \cite{hansson}.

After the projection, the two quasiholes should also be identified,
$H_+\sim J^+_0 H_-$; moreover, they should occur 
in even number with vanishing total isospin.
For four quasiholes, there are three possible amplitudes
that are associated to the same physical excitation;
the first is:
\ba
\Psi_{(12,34)}& = &{\cal A}_{z_i}\Big\langle
\left[ H_+(\eta_1) H_+(\eta_2) H_-(\eta_3) H_-(\eta_4) 
\right.
\nl
&& \left.\left.
\qquad +\  (+ \leftrightarrow -)\right] \prod V_e(z_i) \right\rangle,
\label{4qh}\ea
where we collectively denoted by $V_e$ the two electron fields.
The other two amplitudes, $\Psi_{(13,24)}$ and $\Psi_{(14,24)}$,
amount to different assignments of $(\pm)$.
These three amplitudes give rise to a multidimensional representation
of the braid group ${\cal B}_4$, i.e. to non-Abelian statistics.
 
Therefore, the description of Jain quasiholes within 
the $\winf$ minimal models naturally implies their
non-Abelian fractional statistics. This is our second main result.

Inspection of the $2k$-hole states for $k=2,3$ shows that there are no
degeneracies among the possible amplitudes, in contrast with the
Pfaffian case \cite{nw}; thus, their number grows like
$d_k=2k!/2(k!)^2\sim 2^{2k}/\sqrt{k}$. The same multiplicities can be
inferred by tensoring isospin one-half representations as predicted by
CFT.

The Jain theory also considers quasiholes made in a single effective
level, involving one type of field only, say $H_+$. These excitations
obey Abelian fractional statistics, and are not allowed in the $\winf$
minimal models because they violate the condition of vanishing total
isospin.  Their actual relevance depends on the size of their gap, not
predicted by the CFT approach.

We remark that the results presented here do not yet provide a proof of
non-Abelian statistics in Jain states, lacking an understanding of the
energetics of excitations. In the case of the Pfaffian state \cite{rr},
the three-body short-distance electron potential
identifies the exact zero-energy subspace in which the non-Abelian
excitations have the desired degeneracy.  Hopefully, corresponding
steps in the hierarchical states could be made by developing the
studies in \cite{gaffnian}, including numerical analyses of the
projection \cite{disk} and entanglement entropy \cite{ivan},
and complete the arguments presented here.
Present experiments testing non-Abelian statistics
\cite{na-interf} \cite{thermo-exp}
could also be tried on Jain states, if higher magnetic
fields ($\nu<2$) can be attained.

\begin{acknowledgments}
We are indebted to G. R. Zemba for collaboration in the early stages
of this work.  We also thank A. Bernevig, M. C. Diamantini,
T.H. Hansson, N. Regnault, I. D. Rodriguez, K. Schoutens,
C. A. Trugenberger and S. Viefers for interesting discussions. The
Galileo Galilei Institute for Theoretical Physics, Florence, and
Nordita, Stockholm, are acknowledged for their kind hospitality.
\end{acknowledgments}

\end{document}